\documentclass[journal]{IEEEtran}

\usepackage{ifpdf}
\usepackage{amsmath,amsfonts,amssymb}
\usepackage{algorithmic}
\usepackage{graphicx,subcaption}\graphicspath{{./figures/}}
\usepackage{enumitem}
\usepackage{siunitx}
\usepackage{lscape}

\usepackage{soul}

\usepackage[square, numbers, sort&compress]{natbib}

\usepackage{blindtext}
\usepackage{leftidx}

\DeclareMathOperator*{\cov}{cov}

\newcommand*\Npdf{\mathcal{N}}
\usepackage[dvipsnames]{xcolor}

\newcommand*\D{\textsc{d}}
\newcommand*\ex{\textsc{ex}}
\newcommand*\fa{\textsc{fa}}
\newcommand*\nt{\textsc{nt}}
\newcommand*\s{\textsc{s}}
\DeclareMathOperator\tracks{tracks}
\newcommand{\Tau}{\mathrm{T}}

\usepackage{xspace}
\newcommand*\eg{\emph{e.g.}\xspace}
\newcommand*\ie{\emph{i.e.}\xspace}

\usepackage{tikz}
\usetikzlibrary{arrows,automata,calc,patterns,positioning}

\begin{document}

\title{Framework for Network-Constrained Tracking of Cyclists and Pedestrians}

\author{Alphonse~Vial, Gustaf~Hendeby, Winnie~Daamen, Bart~van~Arem, and~Serge~Hoogendoorn
		\thanks{A.~Vial, W.~Daamen, B.~van~Arem, and S.~Hoogendoorn are with the Department of Transport and Planning, TU Delft, The Netherlands.}
		\thanks{G.~Hendeby is with the Department of Electrical Engineering, Division of
			Automatic Control, Link\"oping University, Sweden.}}

\maketitle

\begin{abstract}
The increase in perception capabilities of connected mobile sensor platforms (\eg, self-driving vehicles, drones, and robots) leads to an extensive surge of sensed features at various temporal and spatial scales.  Beyond their traditional use for safe operation, available observations could enable to see how and where people move on sidewalks and cycle paths, to eventually obtain a complete microscopic and macroscopic picture of the traffic flows in a larger area.  This paper proposes a new method for advanced traffic applications, tracking an unknown and varying number of moving targets (\eg, pedestrians or  cyclists) constrained by a road network, using mobile (\eg, vehicles) spatially distributed sensor platforms.  The key contribution in this paper is to introduce the concept of network bound targets into the multi-target tracking problem, and hence to derive a \emph{network-constrained multi-hypotheses tracker} (NC-MHT) to fully utilize the available road information.  This is done by introducing a target representation, comprising a traditional target tracking representation and a discrete component placing the target on a given segment in the network.  A simulation study shows that the method performs well in comparison to the standard MHT filter in free space. Results particularly highlight network-constraint effects for more efficient target predictions over extended periods of time, and in the simplification of the measurement association process, as compared to not utilizing a network structure. This theoretical work also directs attention to latent privacy concerns for potential applications.
\end{abstract}

\begin{IEEEkeywords}
pedestrians, cyclists, trajectory~reconstruction, multiple~target~tracking, pedestrian~tracking, cyclist~tracking, road~network, road~information, moving~sensors, data~association, multiple~hypothesis~tracking, network-constrained~multi-hypotheses~tracker, NC-MHT, traffic~data, traffic~monitoring~and~control
\end{IEEEkeywords}

\section{Introduction} \label{section: introduction}
Inferring the number of pedestrians and cyclists, as well as their individual states (\eg, position, velocity) from a sequence of measurements, allows drawing a complete micro- and macroscopic picture of the traffic flows in an observed area. This is of great value for advanced surveillance applications in traffic operation, control and management.

Traditional setups for these applications rely on stationary or participatory sampling technologies. Stationary sensing systems gather data that can be used to reconstruct trajectories or counts, but only at local scale (\eg, cross- or very short road sections). At the same time, crowd-sourced data from mobile and other wearable devices allow tracking individuals through a network, but require direct or indirect collaboration of the tracked individual and are mostly sparse in nature (not all persons are detected).  Current advanced surveillance applications and traffic monitoring typically consider longer time horizons to associate new measurements with existing tracks to monitor individuals over time.  Proposed methods mainly consider stationary sensor settings, which often produce a visual stream from a fixed location \cite{Barthelemy_2019, Duives_enhancing_2020}, however come with low spatial resolution and scalability.

\begin{figure}[t]
	\centering
	\includegraphics[width=\columnwidth]{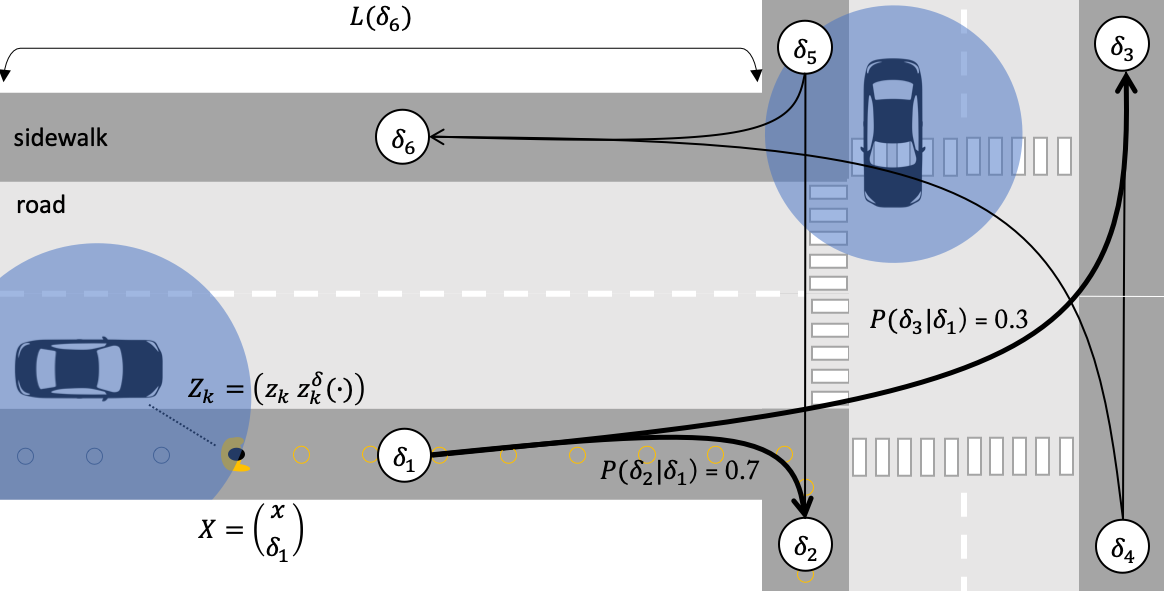}
	
	\caption{Illustration of the concept showing a simple road network and the underlying Markov Process.}
	\label{figure: roadnetwork}
\end{figure}

With novel distributed-computing and connectivity capabilities, a growing number of on-board sensors, the collection of data on pedestrians and cyclists is expected to further intensify with the advent of connected autonomous mobile systems. Self-driving vehicles, drones, or other types of connected robots will enter populated environments and may act as mobile sensing platforms generating a proliferating amount of data about the platform's internal state, but also about the static and dynamic local area they observe. With the collective intelligence and innate mobility of such sensor platforms, pedestrians and cyclists traffic characteristics could be captured at an extended spatial and temporal scale \cite{Vial_2020}. In the future, knowledge about position, motion state, and pose of people could enable next generation traffic or crowd surveillance systems to estimate the number of people and reconstruct trajectories across the network. 

Yet deriving complete trajectories of indistinguishable pedestrians and cyclists in a network using observations from mobile spatially distributed sensor platforms, is complicated. First, there is an unknown and varying number of people, where position and motion states of individuals are unknown. At the same time, noisy sensors, changing environmental conditions, or occlusion are responsible for missed detections and clutter. And with no \emph{a priori} information about which observations originate from which existing or newly detected individual, the many possibilities of assigning a measurement to an individual complicate the task. 

There have been major developments in pedestrian and cyclist detection and tracking, for instance, in areas related to autonomous navigation and control. Safety-critical applications, \eg, self-driving vehicles rely on accurate human motion prediction and path planning \cite{Rudenko_2020, Keller_2014, Kooij_2019}. Proposed methods and pilot deployments however typically consider spatially restricted short time horizons. Thereby presence and state information is crucial to better understand and anticipate actions, that is knowing what an individual will do next, \eg, start, continue, or stop walking. 

In the last decades, different target tracking approaches have been proposed that aim at estimating the number and states of (multiple) dynamic objects using noisy sensor measurements, taking source from the tracking community. Existing algorithms, however, consider short time horizons and are mostly designed for unconstrained motion in two- and three-dimensional space, where complexity tends to grow exponentially due to the number of hypotheses.  Attempts have been made to better model the behavior of pedestrians, \eg, \cite{Batkovic_2018}, which aims at learning to predict for collision avoidance, and \cite{Luber_2011}, in which spatial motion model is learned from observations.  Due to their added complexity, these methods fit poorly into standard multi-target tracking methods.
 
Road information has previously been used to support ground target tracking.  The approach proposed in \cite{pannetierBNR:2004} tracks the objects in free space and then, in a wide sense, projects the estimates onto the road network.  This ensures that targets are estimated on the road, but does not improve projections as done in this paper.  In \cite{Kirubarajan_2000}, a \emph{variable structure interacting multiple model} (VS-IMM) is introduced, which uses different motion models based on the target location to keep the objects on the roads.  In \cite{Ulmke_2006}, targets are described in a combination of global 2-D coordinates, and the quasi 1-D coordinates as used in this paper.  This allows for more efficient predictions while, as in \cite{Kirubarajan_2000, Song_multi-vehicle_2018}, the VS-IMM is used.  Compared to the proposed method, the number of target hypotheses are regularly reduced, keeping them alive until observations can determine which one is correct in the proposed \emph{multi-hypotheses tracker} (MHT).  Furthermore, \cite{Kirubarajan_2000, Ulmke_2006} consider airborne ground moving indicators as sensor, which is different from ground based sensors moving on the road network as introduced in this work.  In more recent efforts, \cite{Song_multi-vehicle_2018} uses measurements from static sensors and include road map information, however, focus on tracking interacting vehicles rather than on the effects of road constraints when tracking targets in a network.  \cite{Zheng_2018} extends the idea of including road map knowledge to random finite set (RFS) methods, however, using a more involved road representation and handling than the method proposed in this paper. \cite{Lopez-araquistain_2019} also uses a different road and coordinate representation that requires projections, however, the concept of branching hypotheses in ambiguous situations (\eg, at intersections), also found in \cite{Krishanth_prediction_2014}, is similar to the handling found in our approach.

The key contribution in this paper is to introduce the concept of network bound targets into the multi-target tracking problem. This is done by introducing a target representation, comprising a traditional target tracking representation and a discrete component placing the target on a given edge in the network.  This model is then used to derive a \emph{network-constrained multi-hypotheses tracker} (NC-MHT). The inclusion of knowledge about the network structure allows for more efficient target predictions over extended periods of time and simplifies the measurement association process, as compared to not utilizing a network structure. The network bound target model is derived with tracking of an unknown and varying number of pedestrians and cyclists in an urban setting using mobile spatially distributed sensor platforms in mind. Hence, the NC-MHT is evaluated on three simulations of targets moving around on an urban infrastructure of connected roads, highlighting different properties introduced by adding the network constraint. The NC-MHT opens up the field for new applications for network-wide traffic surveillance, using information about the number of individuals and their states, to enhance advanced traffic operation, control and management systems. 

The remainder of this paper is organized as follows. The next section presents relevant background theory on fundamentals, namely Bayesian filtering and multiple target tracking. Section~III introduces the mathematical problem formulation and network-constrained system models. Section~IV presents the derived NC-MHT filter and details the integration of the network structure. Section~V presents simulation results from three tracking scenarios. Section~VI discusses the outlook and potential extensions of the proposed approach. Conclusions are drawn in Section~VII.

\section{Background}
Using conventional methods for multiple target tracking (MTT), the individual targets are tracked using Bayesian filters \cite{Fang_2018}, following a step where the available observations in each scan are assigned to the different tracks.  In this context, a track represents a potential target and contains information about its estimated past and current state, and a scan is a set of observations received at the specific point in time. A MHT considers several different association hypotheses in parallel.  Additional logics handle track creation and termination of tracks over time.  The steps are outlined here, and further adapted to network bound targets in the following section.

\subsection{Bayesian Filtering}
The state $x_k$ (\eg, position and velocity) of a target at time $k$ can, given the observations from time $1$ to $k$, $z_{1:k}$, be estimated using recursive Bayesian filtering.  This is achieved using a two step iterative process that sequentially predicts and updates the probability density function of the target state $p(x_k|z_{1:k})$,
\begin{subequations}
    \label{eq: Bayesian Filtering}
    \begin{align}
        p(x_{k+1}|z_{1:k})
        &= \int p(x_{k+1}|x_k)p(x_k|z_{1:k})\,dx_k \\
        p(x_{k}|z_{1:k})
        &= \frac{p(z_k|x_k)p(x_k|z_{1:k-1})}{p(z_k|z_{1:k-1})} .
    \end{align}
\end{subequations}
The first equation predicts future states, given the current state and a motion model $p(x_{k+1}|x_k)$, and the second step incorporates information from a new observation $p(z_k|x_k)$ into the estimate.

For linear models with Gaussian noise, where both the dynamic and measurement model are linear, the  seminal \emph{Kalman filter} (KF) \cite{Kalman_1960} provides the analytic solution to the Bayesian filtering recursion.  For the case that nonlinear models are used, the \emph{extended Kalman filter} (EKF) \cite{Jazwinski_1970}, or \emph{unscented Kalman filter} (UKF) \cite{Julier_1997} can be used to approximate the solution, and in highly nonlinear scenarios the \emph{particle filter} (PF) \cite{Ristic_2004} can be used.

\subsection{Multiple Target Tracking}
Multiple target tracking is an extension to the state estimation problem where both the number of targets and their states should be estimated, based on available scans. The observations that make up the scans are noisy, a target may sometimes fail to produce a detection, and clutter (false observations) which all complicate the task. 

The core of the MTT problem lies in the association of observations to the right track, as given the association, the state of the tracks can be estimated using Bayesian filtering. Data association is needed as it is not known which observation originates from which target, not to mention which  observations are clutter. Classical MTT solutions as the \emph{global nearest neighbor} (GNN) tracker, the \emph{joint probabilistic data association} (JPDA) filter, and the \emph{multiple hypothesis tracker} (MHT) \cite{Reid_1979} differ in the way the association is performed and how many hypotheses are maintained.  See \cite{blackmanP:1999, bar-shalomL:1993} for more details. More recently developed MTT methods, \eg, the \emph{probabilistic hypothesis density} (PHD) filter \cite{Vo_2006, Krishanth_2017}, the \emph{labeled multi-Bernoulli} (LMB) filter \cite{Reuter_2014}, and  \emph{Poisson multi-Bernoulli mixture} (PMBM) filter \cite{garciafernadezWGS:2018}, are based on random finite set statistics and are derived differently, resulting in slightly different methods. None of these methods, however, are designed to utilize network constraints.

The MHT is one of the most popular and common MTT methods. The use of several association hypotheses in parallel took off with Reid's seminal paper \cite{Reid_1979}, yet requires many approximations to become a computationally tractable technique as the number of possible track associations increases exponentially with time. Since then, it has been further developed as this technique gained some momentum in the tracking community, and has been applied in specific application domains. The MHT technique relies on evaluating the probabilities of sequences of measurements from various targets. Two main variants of MHT can be found in literature: Hypothesis-Oriented (HO-) \cite{Reid_1979}, and Track-Oriented MHT (TO-MHT) \cite{Kurien_1990}. Cox \cite{Cox_1996} further contributed to the design of an efficient MHT implementation by introducing Murty's algorithm \cite{Murty_1968}. The latter reduces the computational complexity as it helps avoiding considering unnecessary hypotheses.
\section{Problem Formulation} \label{section: problem forumulation}
This paper considers a regular MTT problem, with an additional network constraint as imposed, \eg, by targets being bound to a system of roads. That is, given a set of mobile sensors, with limited field of view, determine the number of targets present in the tracking volume and their respective state.  Standard target tracking assumptions are assumed to apply:
\begin{itemize}
  \item[A1:] Targets act independently, without influencing one another.
  \item[A2:] The number of new targets at each time is Poisson distributed, with the rate $\lambda_{\nt}$, and targets appear uniformly in the tracking volume.
  \item[A3:] In a scan of observations from a sensor, each observation either originates from a single target or is a false detection, and each target produces at most one observation.
  \item[A4:] The number of false observations (observations not originating from a target) in a scan is Poisson distributed with the rate $\lambda_{\fa}$, and the observations are uniformly distributed in the sensors field of view.
  \item[A5:] The probability to detect a target that is within the field of a sensor is constant, $P_{\D}$.
  \item[A6:] The probability a target survives from a time to the next is constant, $P_{\s}$.
\end{itemize}
Additionally, \emph{this paper assumes targets to be network constrained}, that is:
\begin{itemize}
  \item[A7:] The targets are bound to network constraints, and hence cannot move freely in the tracking volume. A target is always associated with exactly one discrete network state at any time.
\end{itemize}
Here, it will be assumed that the network constraint is in fact a road network.  However, the theory developed applies to any situation where the targets are associated to a discrete state that can be modeled as a Markov process. Notation will be kept as general as possible.

\subsection{The Network Constraint: Road Network}
\label{sec:network-constraint}
Targets are constrained to move on roads, sidewalks, and cycle paths, which make up the considered network constraints defined in Assumption~A7.  A key observation is that a target can only be on one road segment at the time, and once the target reaches an intersection it continues on to another connected road segment, which limits the motion of the targets.  This can be modelled as a discrete Markov chain, where the road segments make up the states and the transition probabilities\footnote{A discussion on the determination of the transition probabilities is provided in Section \ref{subsection: sensing infrastructure and measurements}.} model the probabilities to transition from one road to another once the target reaches an intersection.  This can be represented with a directed graph $G = (V, A)$, with a set of vertices $V$ (representing the different states or road segments) of size $n$ and edges $A \subseteq V \times V$ (the possible transition from one road segment to another) of size $m$.  A weight function $p:A\to [0, 1]$ assigns a non-negative probability $p(\delta'|\delta)$ to transition from state $\delta$ to segment $\delta'$ once the target reaches the end of the road segment represented by $\delta$.  It follows from the rule of total probability that $\sum_{\delta'} p(\delta'|\delta)=1$.  Furthermore, each vertex contains information about the length of the road segment it represents, $L(\delta)>0$.  The graph hence represents a topological map of the road network.

\subsection{Network-Constrained State-Space Model}
\label{sec:NCSSM}

To utilize the road constraints, a discrete component is added to the target state to indicate which road segment the target is on, and the ``regular'' part of the state will be used to track the target within the node.  Both the prediction and observation model are modified to take the discrete component in consideration.

Targets evolve independently of each other, in discrete time, and the state of each target is represented by a vector
\begin{equation*}
    X_k = \begin{pmatrix} x_k \\ \delta_k \end{pmatrix} ,
\end{equation*}
where $x_k$ is a real-valued state vector describing the target motion within a node, and $\delta_k$ is the state in the Markov process. The semantics of the state space $x_k \in \mathbb{R}^{n_x}$ can vary depending on the considered application, yet usually includes components related to the target's kinematic state such as position, velocity, and possibly acceleration, along the road segment, but could also be used to describe the lateral position on the road. In the end, the actual position of the target in the world is a combination of the current node $\delta_k$, giving the general position determined from map, and the $x_k$ component which specifies the position along the road segment as illustrated in Fig.~\ref{figure: roadnetwork}.

\subsubsection{Motion Model}

The nonlinear dynamic model describes the motion within a node as
\begin{equation}
    x_{k+1}
    = f(x_{k}) + v_{k}, \quad v_{k} \sim \mathcal{N}(0,Q_{k}) 
\end{equation}
where $v_k$ is process noise that is assumed to be white and Gaussian with covariance matrix $\cov{v}_t = Q_k$.  The discrete state $\delta_k$ follows the transition probability $p(\delta_{k+1}|\delta_k, x_{k+1})$ where $\delta_{k+1}=\delta_k$ unless $x_{k+1}$ indicates that the target reached the end of the current road segment.  Once a state transition occurs, the node specific state, and distance traveled along the current road segment, $x_{k+1}$ is updated accordingly. Typically, this means resetting the distance traveled.  In the considered application, when a transition takes place and $x^p_{k+1}$ is the position component of the continuous state, remove the distance traveled on the previous segment to reach the end of it  $x^p_{k+1}\leftarrow x^p_{k+1} - L(\delta_k)$.  It could, however, potentially also include additional adjustments as a result of specific properties of the new road segment.  The adjustment procedure must be repeated if after compensation the target is past the end of the next segment too.

\subsubsection{Observation Model}

At any time instant $k$, sensor platforms may deliver scans, $\mathcal{Z}_k$. The position of the mobile sensor is assumed to be known, and as defined in Assumption~A5, each target is detected with probability $P_{\D}$ when in the field of view of a sensor. The state variables above are related to measurements $Z_k$ according to a nonlinear observation model 
\begin{subequations}
\begin{align}
    Z_k
    &= \bigl(z_k, z^\delta_k(\cdot)\bigr)\\
    z_k
    &=  h(X_k) + e_{k}, &  e_{k} &\sim \mathcal{N}(0,R_{k})\\
    z_k^\delta(\delta')
    &= \Pr(\delta_k = \delta')
\end{align}
\end{subequations}
where $h(x_k)$ represents the state of the target and relates to an ideal (vector valued) measurement, and  $e_{k}$ is the measurement error  which is assumed to be Gaussian with covariance matrix $\cov(e_k) = R_{k}$. Furthermore, each observation includes the probability that the target was observed in the discrete state $\delta'$, $z_k^\delta(\delta')$ --- a simplification, reasonable in many cases, is that the discrete state information is accurately provided, \ie, that only one state has non-zero probability. 

Because several sensors can operate simultaneously, let $S = \{s_v\}_{v=1:V}$ be the set of mobile sensors, where $V$ is the number of operating sensors that acquire data of dynamic targets present on the road network. A scan $\mathcal{Z}^v_k$ is the set of all $M^v_k$ measurements received by a sensor platform $v$ at time $k$ such that $\mathcal{Z}^v_k = \{Z^{v, 1}_k,Z^{v, 2}_k,\dots, Z^{v,M_k}_k\}$, and where $Z^{v,i}_k$ represents the $i$th observation received at scan $k$, by sensor $v$. This way, a target may be observed by multiple sensor platforms at the same time instant.

\subsection{False Observations and New Targets}
As part of the point-to-object and defined in Assumption~A3, each target gives rise to at most one measurement per sensor, at each time step. Targets in the field of view are assumed detected and observed with probability $P_{\D}$ (Assumption~A5). Further, the number of new targets $n_{\nt}$ and false observations $n_{\fa}$ in the volume $V$ are assumed Poisson distributed,
\begin{equation}
    p(n,\lambda, V)
        = e^{-\lambda V}\frac{(\lambda V)^n}{n!},
\end{equation}
with intensity $\lambda_{\nt}$ and $\lambda_{\fa}$, respectively, see Assumptions~A2 and~A4. Both new targets and false observations, together denoted extraneous observations, are assumed uniformly distributed in the current tracking volume. The total spatial density of the extraneous observations also follows a Poisson distribution with intensity $\lambda_{\ex}$ and can be expressed as follows
\begin{equation}
    \lambda_{\ex}
    \triangleq \lambda_{\nt} + \lambda_{\fa},
\end{equation}
which, as will be seen, can be used to simplify some expressions.

\textbf{Note:} In the network constraint cases, the sensors can be assumed to only produce state (road) bound observations, which is beneficial compared to the general case where no such restrictions exist.
\section{Network-Constrained Multiple Hypothesis Tracking} \label{section: mht}

\begin{figure*}[t!]
    \centering
	\includegraphics[width=\textwidth]{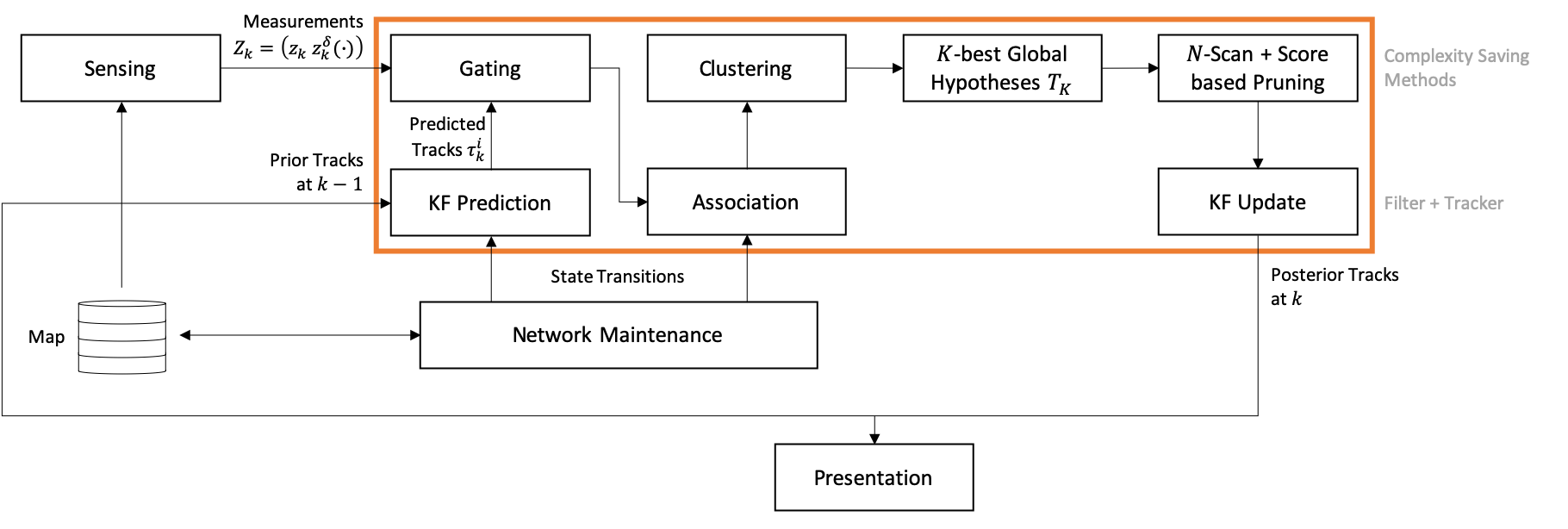}
	\caption{Outline of proposed tracking framework, with traditional MHT logic inside orange area.}
	\label{figure: outline}
\end{figure*}

In a MHT, several potential targets are tracked using separate single target tracking filters, and a higher level logic decides which observations to pair with the currently maintained tracks, which observations should be considered to be false, and when to create new tracks.  In the MHT, several such hypotheses are maintained in parallel, while at the same time estimating the probability of each of the different hypotheses to be correct.  Key components are to generate the appropriate hypotheses, and to reduce the number of considered hypotheses as much as possible to keep the computational complexity as low as possible.  The network constraint, being discrete in its nature, simplifies the hypothesis handling. This section outlines the MHT formulas adopted to take the network constraints into consideration. Fig.~\ref{figure: outline} provides an overview of the different elements of the proposed algorithm.

\subsection{Extended Kalman Filter} \label{subsection: EKF}
The EKF \cite{Jazwinski_1970} is arguably the most common technique used for single target tracking in the MHT.  The EKF does not deal with discrete components, hence it cannot be directly applied to the hybrid state proposed in Sec.~\ref{sec:NCSSM}.

\subsubsection{Time Update}
The solution is to condition on the discrete state $\delta_k$, and use the regular EKF for the continuous part of the state $x_k$.  This yields, for a given $\delta_k$, for the time update
\begin{subequations}
\label{eq:EKF:tup}
\begin{align}
    \label{eq:xhat}
    \hat{x}_{k|k-1}
    &= f(\hat{x}_{k-1|k-1}),\\
    \label{eq:P_{k|k-1}}
    P_{k|k-1}
    &= F_{k}P_{k-1|k-1}F_{k}^T + Q_{k}.
\end{align}
\end{subequations}
where $\hat x_{k|k-1}$ is the mean of the estimate of the state at time $k$ given the measurements up until time $k-1$, $P_{k|k-1}$ is the matching error covariance matrix, and $F_k = f'(\hat{x}_{k-1|k-1})$ the linearization around $\hat{x}_{k-1|k-1}$ of the dynamic model.

Caveat, performing this prediction step might position the target past the end of the current discrete state.  The interpretation in the described setting is that the target reaches the end of the current road segment.  That is, the target should transition to a new discrete state.  This is done following the underlying Markov model as described in Section~\ref{sec:network-constraint}.  
If there are several different possible transitions, where each have transition probabilities, all are taken and new track hypotheses are created, and the hypothesis probability is updated accordingly by the MHT logics, as described below.  After the transition the continuous state $x$ is compensated in each transition, as described in Sec.~\ref{sec:NCSSM}.  As for the case there is no discrete state to be propagated to, \eg,  a target is existing the tracking volume of interest, the target is simply removed from the set. 

\subsubsection{Measurement Update}
When an observation is obtained, again conditioned on the discrete state, the continuous part of the state can be updated, provided the discrete state is possible in the observation, using
\begin{subequations}
\label{eq:EKF:mup}
\begin{align}
    \label{eqn: kf Xkk}
    \hat{x}_{k|k}
    &= \hat{X}_{k|k-1} + K_k(z_{k}-\hat z_k), \\
    \label{eqn: kf Pkk}
    P_{k|k}
    &= P_{k|k-1} - K_{k}H_{k}P_{k|k-1},\\
    \hat Z_k
    &= h(\hat{x}_{k|k-1})\\
    \label{eqn: kf S}
    S_k &= H_{k}P_{k|k-1}H_{k}^T + R_{k},\\
    \label{eqn: kf K}
    K_{k} &= P_{k|k-1}H_{k}^TS_{k}^{-1},
\end{align}
\end{subequations}
where $H_K = h'(\hat{x}_{k|k-1})$.  If no observation is available, the update step is simply skipped, and if several are available the step is repeated.

Contrary to the prediction step, the observation update will never trigger transition in the discrete state, if the simplified version is used. The update is performed under the presumption that the observation places the target in the current discrete state, hence the estimate stays in the same state.  The only ambiguities in this step is due to different observations being associated with the current track.

\subsubsection{Corner Cases}

The above approach assumes that each track hypothesis can fully be assigned to a discrete state.  This is often a reasonable approximation as long as the estimated track is far from the end of the current road segment, as compared to the uncertainty of the estimate.  However, if the track is close to the edge of the road segment or the uncertainty is considerable, with a non-negligible probability the track could instead be on an adjacent road segment.  Two different solutions to this are considered here: always splitting tracks with significant overlap, and splitting when needed for proper association with observations.

In the former case, when a track is predicted close to an intersection, split the track in $n-1$, where $n$ is the number of potential roads at this intersection, and divide the probability according to the probability mass on all segments other than the segment the target is currently on (based on the uncertainty).  The resulting two new tracks should have means combining to the original mean, and a combined covariance (including the spread of the mean term) should match the original covariance.  A benefit of this approach is that it is easy to implement, however, it leads to unnecessarily fast growth of hypotheses to consider.

In the latter case, observations are allowed to be associated with tracks on adjacent discrete states, if there is a significant probability of leakage.  Associated this way, the track is split into two identical ones splitting the probability as above, and where the one is projected back at the extension of the adjacent segment (\ie, as if a transition had taken place but $x_k$ had not yet been adjusted) and then that track is updated as if it had been on the segment.  This results in fewer hypotheses, as splitting is only performed when needed, but the logic is slightly more complex to implement.

\subsubsection{Track Representation}

As should be clear from the presentation above, a specific estimated track, $i$, is fully described by the sequence of discrete state transitions assumed to take place, $\delta^i_{1:k}$, and the observations associated with the track, $\theta^i_{1:l}$, together denoted $\tau^i_{k|l}= (\delta^i_{1:k} ,\theta^i_{1:l})$, $k\ge l$ for the state at time $k$ with measurements from up until time $l$.  And several different $\tau_{k|l}$ could describe the same underlying target, with different assumptions about the discrete state sequence and what observations originate from the target.  

New tracks are started where available measurements are not associated with any other existing track.  The observation is used to set the general position determined from the map and the position on the road segment such that given $Z_k$,
\begin{subequations}
		\begin{align}
		x_k
		&= h^{-1}(z_k) \\
		\delta_k
		&= z_k^\delta(\delta') = \Pr(\delta_k = \delta').
		\end{align}
\end{subequations}
If the state $x_k$ cannot be uniquely determined from $z_k$ ($h$ is not invertible), as much as possible is derived from $z_k$ and remaining parts of the state vector is set to default values.  As an example, assume $x_k$ state with position and velocity and measurements $z_k$ providing (possibly via a conversion) the position; then the position part of $x_k$ is given by $z_k$ and the speed is set to a nominal value, \eg $0$.  This represents the possibility that the observation corresponds to a new target that has entered or newly appeared in the network.

\begin{figure*}[ht!]
    \centering
	\includegraphics[width=\textwidth]{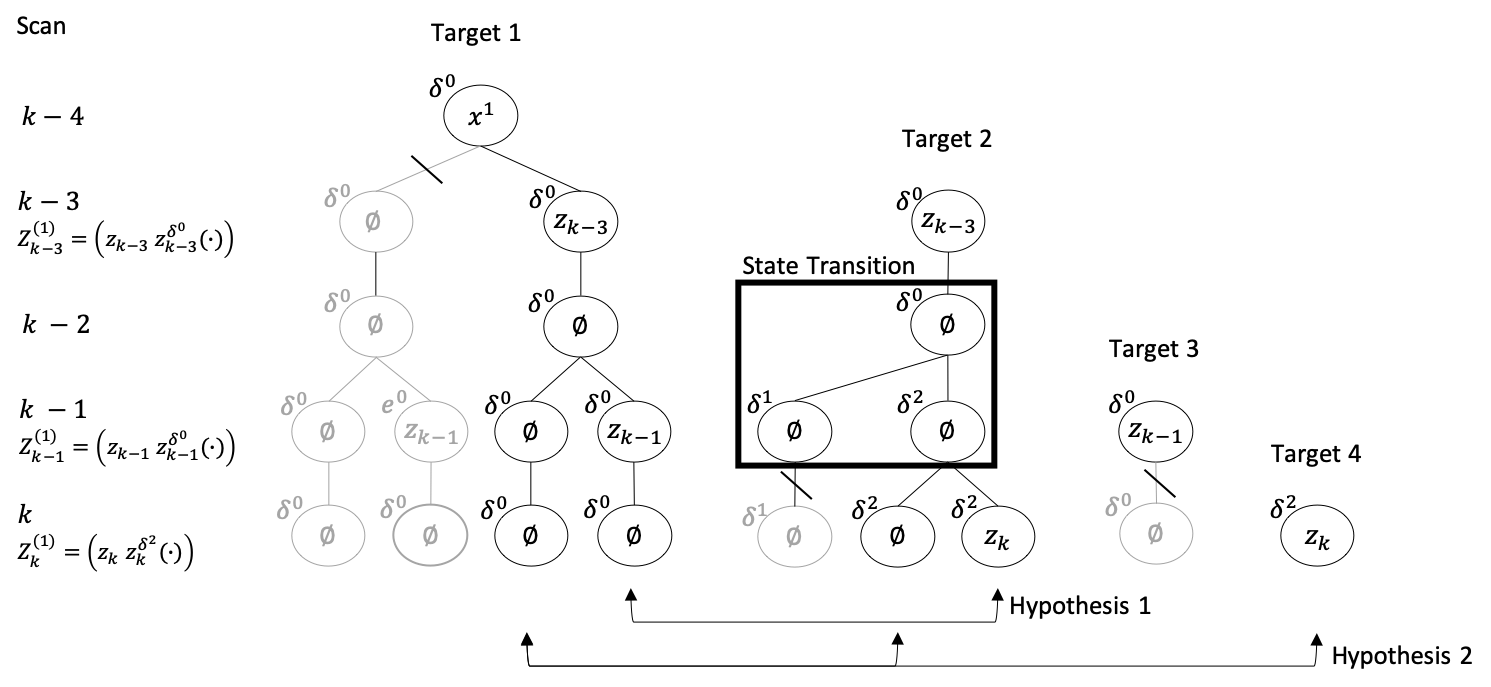}
	
	\caption{Illustration of track hypotheses and global hypotheses. A target tree can be constructed to store the hypothetical tracks of the corresponding target. A global hypothesis represents a set of all possible tracks that are compatible from different targets. The state transition occurs when $x_{k+1}$ indicates that the target reached the end road segment, here indicated by the state $\delta^0$.}
	\label{figure: hypotheses}
\end{figure*}

\subsection{Track Hypotheses and Scores}
\label{sec: track hypotheses and track scores}

A track comprises the different hypotheses about a target, as a result of different $\tau_{k|l}$ assumptions as described above, and a score for each of these.  The different track hypotheses assumed to represent the same underlying target can be represented with a track tree, where the tree branches are a result of discrete state transitions and how the observations are associated.  The score indicates the importance of the specific hypothesis.  For track hypothesis $i$ this is described as the log-likelihood ratio \cite{Bar-Shalom_2007}
\begin{equation}
    \ell^i_{k|l}
    = \log \frac{p(\text{target exist}|\tau^i_{k|l}, Z_{1:l})}{p(\text{no target}|\tau^i_{k|l}, Z_{1:l})}.
\end{equation}
The score can be updated recursively, as decision points are reached in the track tree.

Similar to the track state, the track score update can be divided into two steps.  When the track is propagated in time, except for compensating for the probability to survive affecting all tracks alike, only the changes of discrete states affect the score when the score is divided amongst the potential transitions.  The score of track hypothesis $i$ becomes
\begin{equation}
    \ell^i_{k|k-1}
    = \ell^{p(i)}_{k-1|k-1} + \log\bigl(P_{\s} p(\delta_k|\delta_{k-1}, \hat x^i_{k|k-1})\bigr) ,
\end{equation}
where $p(i)$ is the parent hypothesis (\ie, a single prior hypotheses) of hypothesis $i$.  It should be noted that, when no state transition is possible $\log\bigl(p(\delta_k|\delta_{k-1}, x^i_{k|k-1})\bigr)=\log(P_{\s})$.  If $P_{\s}=1$ this means the score remains unchanged.

The update of the score as the result of assigning a new observation (or not) as the result of obtaining a scan is a bit more involved,
\begin{equation}
    \ell^i_{k|k}
    = \ell^{p(i)}_{k|k-1} +
    \begin{cases}
        \log(1-P_D),
        & \text{if $\theta^i_k = \emptyset$}\\
        \log\Bigl(\frac{P_{\D} p(Z_k|Z_{1:k-1}, \tau^i_{k|k})}{\lambda_{\fa}}\Bigr), & \text{if $\theta^i_k\neq \emptyset$}
    \end{cases},
\end{equation}
where
\begin{equation}
\begin{split}
    p(Z_k|Z_{1:k-1}, \tau^i_{k|k}) & = p(z_k|Z_{1:k-1}, \tau^i_{k|k})\cdot z^\delta_k(\delta_k) \\
    &= \Npdf(z_k|\hat z^i_k, S^i_k)\cdot z^\delta_k(\delta_k) .
\end{split}
\end{equation}
There are two different cases: an observation is obtained or not.

If no observation is obtained, $\theta^i_k=\emptyset$, the only change to the score is that the probability of the target is scaled by the probability to not observe the target, $(1-P_{\D})$.  That is, the score drops if the target is expected to have been observed.

If an observation is obtained, $\theta^i_k\neq \emptyset$, the probability of the target is compensated by the probability to observe the target, $P_{\D}$, and the probability that the target would produce the obtained observation (which can be computed from the innovation in the Kalman filter).  The no target probability is scaled with the probability of obtaining the observation without a target present, $\lambda_{\fa}$.

\textbf{Note:} The absence of observations, sometimes denoted negative information, represents a valuable source of  information as it lowers the score for track hypotheses where the target was not observed as expected.  Hence, though the track state does not change, it is important to update the track scores based on this negative information.  This helps to reduce the number of track hypotheses that should be considered.

New tracks are created from observations and given the initial score $\ell^i_{k|k} = \log(\gamma_{\nt})$, where $\gamma_{\nt}$ is used to get the right initial likelihood ratio.  Terminating tracks, on the other hand, are simply dropped by cause of being too unlikely.

Given the track score $\ell^i_{k|l}$, the probability of track $i$ existing can be computed as
\begin{equation}    
    \label{hypothesis likelihood NLL}
    p(\text{target $i$ exists}|Z_{1:l}, \tau^i_{k|l})
    = \frac{e^{\ell^i_{k|l}}}{1+e^{\ell^i_{k|l}}},
\end{equation}
where it has been utilized that the probability of existence and non-existence must sum to $1$.

\subsection{Global Hypotheses}
A global tracking hypothesis, $\Tau_{1:k}$, is a description of a complete solution to the multi-target tracking problem; \ie, the number of targets and their respective states.  A global hypothesis assigns each observation to a track, or assumes it is a false observation, and determines the sequence of discrete states of all tracks.  This can also be described as picking out which track hypotheses to use.  The number of track hypotheses grows rapidly over time, as ideally all possible discrete state transitions and measurement associations should be explored.  However, only a few of these track hypotheses are consistent with each other, fulfilling Assumption A3 which states that a measurement can only originate from one target, and can form global hypotheses.  A global hypothesis, therefore, contains at most one hypothetical track from each track tree.   Fig.~\ref{figure: hypotheses} illustrates track trees, track hypotheses and global hypotheses, and how they are updated as new scans of observations arrive.

The probability of a global hypothesis is obtained as the product of the probability of each included track hypothesis and the probability of all unassociated measurements being false alarms.  The log-probability of the global hypothesis $l$ can be computed as the sum of the score of the contained track hypothesis compensated with a term taking false observations into consideration
\begin{equation}
    \log\bigl(p(\Tau_{1:k}^l|Z_{1:k})\bigr)
    = \!\!\sum_{i\in\tracks(l)}\!\! \ell^i_{k|k}+|Z_{1:k}|\log(\lambda_{\fa}) ,
\end{equation}
where $\tracks(l)$ is the set of track hypotheses in global hypothesis $l$, and $\lvert Z_{1:k}\rvert$ the total number of observations.

The global track score can be updated recursively, using the recursive property of the track hypotheses scores
\begin{multline}
\label{eq: hypothesis likelihood}
    \log\bigl(p(\Tau_{1:k}^l|Z_{1:k})\bigr)
    = \log\bigl(p(\Tau_{1:k-1}^l|Z_{1:k-1})\bigr) +\\ 
    \sum_{i\in\tracks(l)} \ell^i_k+|Z_{k}|\log(\lambda_{\fa}) ,
\end{multline}
where $\ell^i_k = (\ell^i_{k|k}-\ell^{p(i)}_{k-1|k-1})$ is the score increment in the current step for track hypothesis $i$.

In this formulation, the network constraints are naturally handled, as they are contained in the track scores, and do not require any special handling.

\subsection{Complexity Saving Methods}

To achieve a computationally tractable solution, it is necessary to lower the number of considered global hypotheses.  This is done by removing too improbable hypotheses, and if possible only generate relevant hypotheses. The methods used for this are discussed in this section.

\subsubsection{Generating the $k$-Best Hypotheses}

The number of global hypotheses grows exponentially with time, and needs to be handled to achieve computational tractability.  One way to limit the complexity is to only generate a subset of the possible hypotheses.  That is, to construct a subset of data associations $\tilde{\Theta}_k \in \Theta_k$ possible in each step, such that $\lvert \tilde{\Theta}_k\rvert \ll \lvert\Theta_k\rvert$, which will generate the  best scores.   This can be posed as an assignment problem and solved with standard combinatorial optimization algorithms.  Many algorithms exist to efficiently solve the optimal assignment problem based on the additive increments in the track scores, \eg, the Hungarian algorithm \cite{Kuhn_1955}, the Auction algorithm \cite{Bertsekas_1988}, and the Jonker-Volgenant-Castanon \cite{Jonker_1987} algorithm.  If combined with Murty's algorithm \cite{Murty_1968}, the $k$-best assignments can be found in polynomial time.  The remaining hypotheses can then simply be left unconsidered.

\subsubsection{Gating}

Gating is a technique to simplify the association problem, making it less computationally demanding by removing potential associations that are too unlikely.  This is achieved by ignoring track to observation associations, if the observation lies outside the gate of the track.

The gate is set such that the probability of rejecting a true association is low.  Using ellipsoidal gating \cite{Collins_1992} the gate is defined by
\begin{equation}
    D_{\textsc{g}}
    = (z_k-\hat{z}_{k|k-1})^TS^{-1}_k(z_k-\hat{z}_{k|k-1})
    \leq \gamma_{\textsc{g}} ,
\end{equation}
which defines an ellipsoid around the predicted observation in which the observation must be considered.  Assuming Gaussian distributions, 
$D_{\textsc{g}} \sim \chi^2(n_z)$, it is possible to design $\gamma_{\textsc{g}}$ for a given probability to reject a correct association. 

\begin{figure}
    \centering
	\includegraphics[width=\columnwidth]{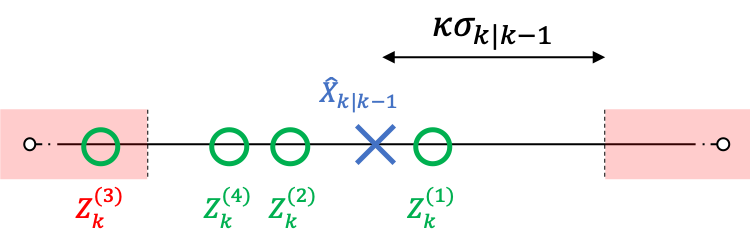}

	\caption{Road-constrained gating.}
	\label{figure: gating}
\end{figure}

With the hybrid state space, as considered here, gating is performed in two steps, first observations where the discrete state does not match are dropped, and then normal gating is performed.  Fig.~\ref{figure: gating} exemplifies the gating procedure, which is simplified by the fact that the continuous part of the observation is scalar such that
\begin{equation}
    |z_k-\hat{z}_{k|k-1}|
    \leq \kappa\sigma_{k|k-1},
\end{equation}
where $\kappa$ is a factor times standard deviations, $\sigma_{k|k-1}$ in the considered dimension.

\subsubsection{Clustering}
\label{clustering}

Clustering is a method in which tracks that share observations are put in clusters, and then each cluster is treated independently.  This is beneficial as association and track hypotheses generation scales poorly with the number of involved targets and observations.  When producing the final global hypotheses, hypotheses from the different clusters can be combined freely.  The clustering is performed based on which tracks allow the same observation in their gate.

Clusters can be formed recursively over time.  In that case, two clusters that gate in the same observation must be merged.  Similarly, clusters should regularly be examined to see if they can be split, as a result of removing track hypotheses that have connected sub-clusters.  For more details  and implementation details the reader is referred to \cite{Reid_1979,Kurien_1990,Olofsson_2017}.  Once again, the hybrid nature of the suggested model can help simplify the handling.

\subsubsection{Pruning Strategies: Track, Hypothesis and Target Management}

Tracks are deleted based on the track score \cite{Blackman_2004}, or $N$-scan sliding window \cite{Kim_2015}.  The first method prunes low probability hypotheses and tracks.  $N$-scan pruning, on the other hand, traces back to the node at scan $k-N$ and deletes the subtrees that diverge from the selected branch at that node.  This represents the possibility that a target has disappeared from the scene.  Furthermore, as indicated in Section \ref{subsection: EKF}, the network constraints cause tracks that can not be propagated to a new discrete state to be removed from the set of existing tracks.  This represents the possibility that a target has exited the network.  Targets are also removed accordingly, that is as soon as it has no remaining track in any hypothesis.  Because the track score contain all relevant statistical information, remaining tracks lose no information during the pruning process.

\section{Simulation Study} \label{section: simulation study}
In this section, the results from an empirical evaluation on three simulated scenarios are presented.  The scenarios are chosen to highlight properties of the proposed NC-MHT on track-handling at intersections, data association, and measurement dependency.  To evaluate the proposed approach, the NC-MHT filter was integrated into the standard (unconstrained) MHT filter from \cite{Olofsson_2017}, to which it is compared\footnote{The compared trackers are implemented in Python and C++, and the simulations were run on a laptop with a 2.8\,GHz Quad-Core Intel Core i7 and 16\,GB memory.} in terms of tracking performance.  Note that it is difficult to provide any accurate complexity analysis, and that the proposed method has low maintenance to keep branching tracks alive, that is, carries more hypotheses that can be easily handled.  The comparison is performed in terms of summarizing statistics of the trackers and the MTT measure \emph{generalized optimal sub-pattern assignment} (GOSPA) metric  \cite{Rahmathullah_2017},
\begin{equation}
\label{eq: GOSPA}
 d_p^{(c,2)}(\textbf{x},\hat{\textbf{x}})
 \!=\!\biggl(\min_{\gamma \in \Gamma}\!\! \sum_{(i,j)\in\gamma}\!\! d(xi, \hat{x}_j)^p + \frac{c^p}{2}(\lvert\textbf{x}\rvert+\lvert\hat{\textbf{x}}\rvert-2\lvert\gamma\rvert)\biggr)^{\frac{1}{p}}
\end{equation}
where $\hat{\textbf{x}} = \{\hat{x}_1,\dots,\hat{x}_{|\hat{\textbf{x}}|}\}$ is a finite subset of state estimates of the ground truth set $\textbf{x} = \{x_1,\dots,x_{|\textbf{x}|}\}$, and $\gamma$ represents the assignment sets between these two sets. Let $d^c(x, \hat{x})=\min(\lVert x-\hat x\rVert, c)$ denote a cutoff metric for any $x, \hat{x} \in \mathbb{R}^n$. Further, let $\Gamma$ denote the set of all possible assignment sets. GOSPA allows a decomposition of the error into three parts: 1) localisation error, 2) missed targets, and 3) false targets.  In the simulations, the GOSPA is computed based on the 2D positions of the tracks, and the parameters are set to $c=8$, corresponding to a maximum error appropriate for this setting, and $p=2$, corresponding to using the 2-norm which is a standard choice. 

Note that GOSPA is a single unified performance metric, whereas the other quantities are presented because they represent properties that are important in MTT.

\begin{figure}[ht]
	\centering
	\begin{subfigure}{\linewidth}
		\centering
		\includegraphics[width=0.85\columnwidth]{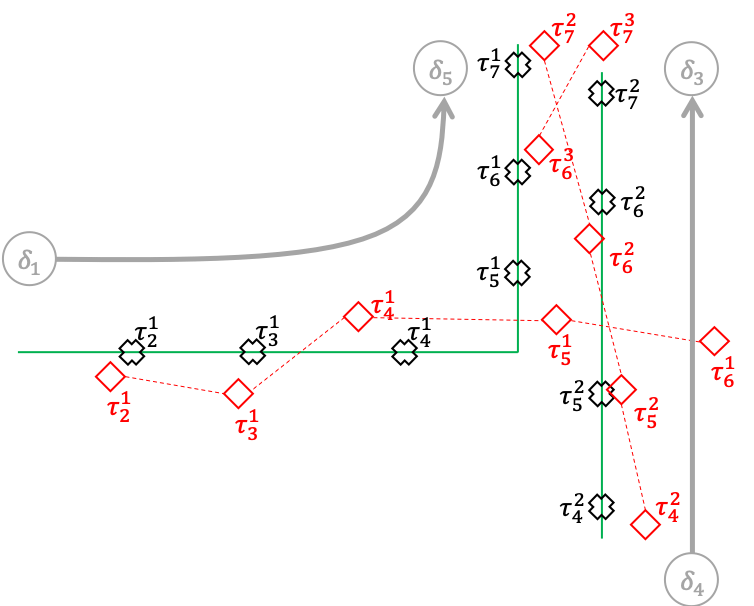}
		\caption{Illustration of NC-MHT and the standard two-dimensional MHT, where ground truth is represented as solid green line.  The inclusion of network constraints allows for more efficient target prediction over longer time periods, and simplifies the measurement association.}
		\label{figure: Illustration tracking behaviour}
	\end{subfigure}
	
	\begin{subfigure}{\linewidth}
		\centering
		\includegraphics[width=\columnwidth, trim = 1.5cm 1.5cm 1.5cm 1.5cm, clip]{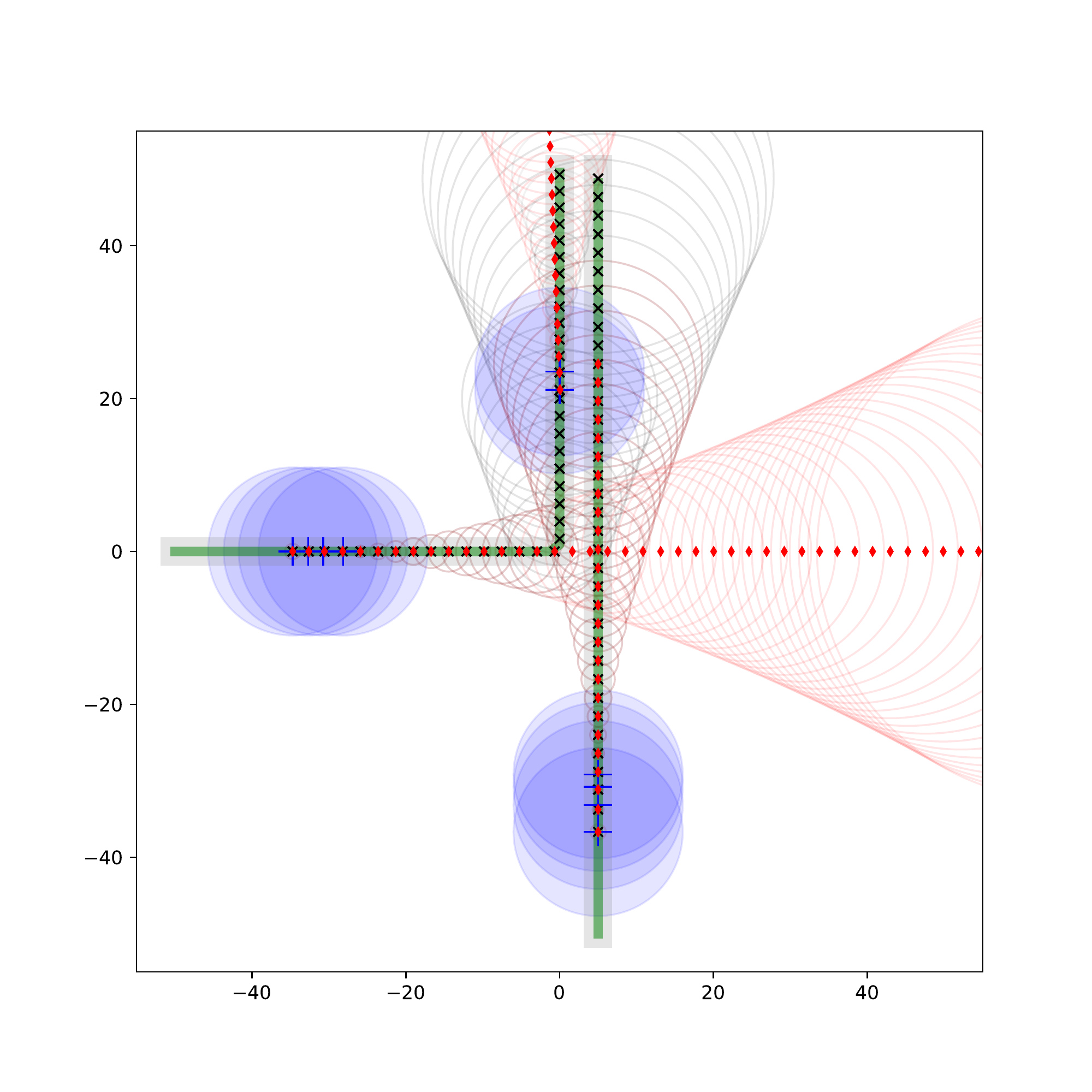}
		\caption{Tracking results at different time steps, visualizing the position and uncertainty of estimated tracks in two dimensions.  Note that \mbox{2-D} allows to visualize the evolution of the [NC-MHT] uncertainty, while it kept only \mbox{1-D} (\ie, along the road) in the implementation. The blue crosses are sensor measurements and are at the center of the sensor's field of view (blue circle).}
		\label{fig: Results tracking behaviour}
	\end{subfigure}  
	\caption{Illustration and simulation results of main differences in tracking behaviour between NC-MHT (black crosses) and the standard MHT (red diamonds).}
	\label{figure: ncmhtvsmht}  
\end{figure}

\subsection{Simulation Setup}
The kinematic state at time $k$ is described by the target's position and velocity.  For simplicity one-dimensional motion is considered for the NC-MHT such that $x_k =  [y_k, \Dot{y}_k]^T$ contains position and velocity.  Nonlinear measurement models are more adequate when using real-world data, as not only position and velocity of a target is measured, but also range $(r)$, bearing $(\alpha)$, and angle rate $(\Dot{r})$ to a target's position can be measured (\eg, radar).  The choice of an appropriate model depends on the considered application, yet this implementation uses linear-Gaussian dynamics and observations as handling the nonlinear case is a trivial extension following \eqref{eq:EKF:tup}--\eqref{eq:EKF:mup}.  The kinematic state motion model follows linear constant velocity (CV) given parameters
\begin{align*}
    F
    &= \begin{bmatrix}
      1 & \Delta_T  \\
      0 & 1
    \end{bmatrix},
    &
    Q
    &= q^2 \cdot \begin{bmatrix}
      \Delta_T^3/3  & \Delta_T^2/2 \\
      \Delta_T^2/2  & \Delta_T
    \end{bmatrix}
\end{align*}
with sampling time $\Delta_T= \SI{1}{\second}$, and where $q=\SI{0.1}{\metre}$ is the process noise variance.  In the simulation, targets move with initial mean speed \mbox{$\SI{1.415}{\metre/\second}$ (standard deviation is $\SI{0.215}{\metre/\second}$)}.  Nonlinear effects in the target dynamics at intersections are ignored in this simulation, but can easily be considered using models describing interactions between, \eg, targets at higher density junctions. For simplicity the probability of target survival is set to $P_{\s}=1$. The discrete state transition probabilities, moving targets from segment to segment, are equally distributed among the set of possible outgoing transitions from any given node.

\begin{figure*}[t!]
	\centering
	\begin{subfigure}[t]{0.25\linewidth}
		\centering
		\includegraphics[width=\linewidth]{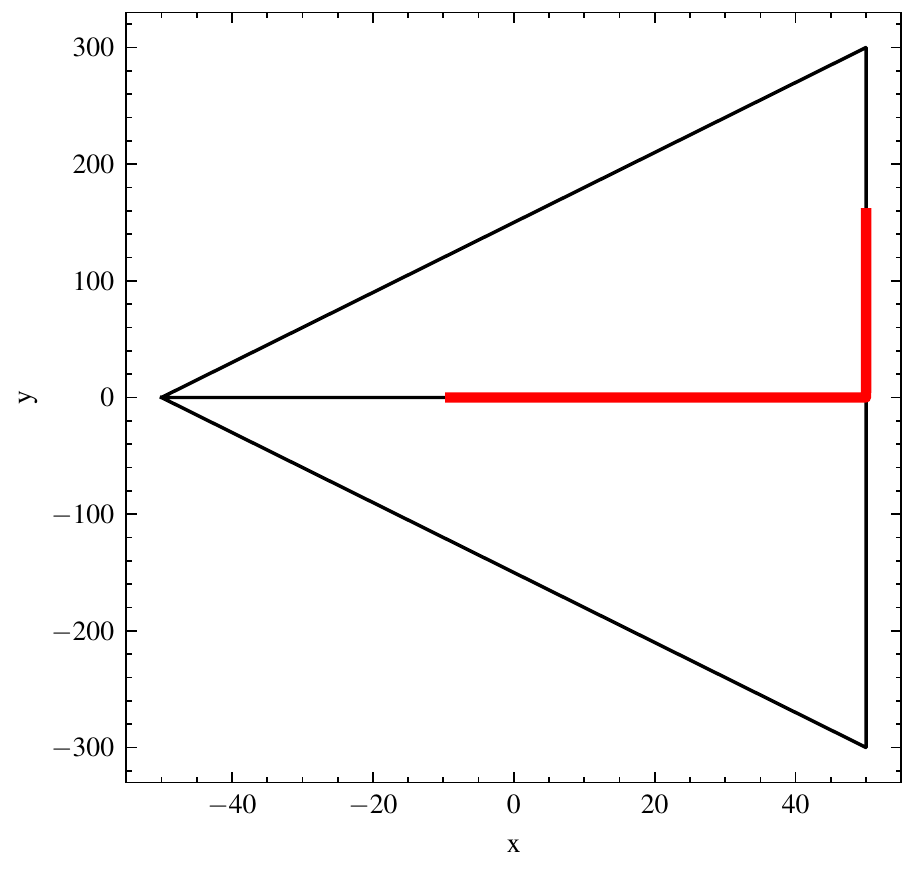}
	\end{subfigure}%
	~ 
	\begin{subfigure}[t]{0.25\linewidth}
		\centering
		\includegraphics[width=\linewidth]{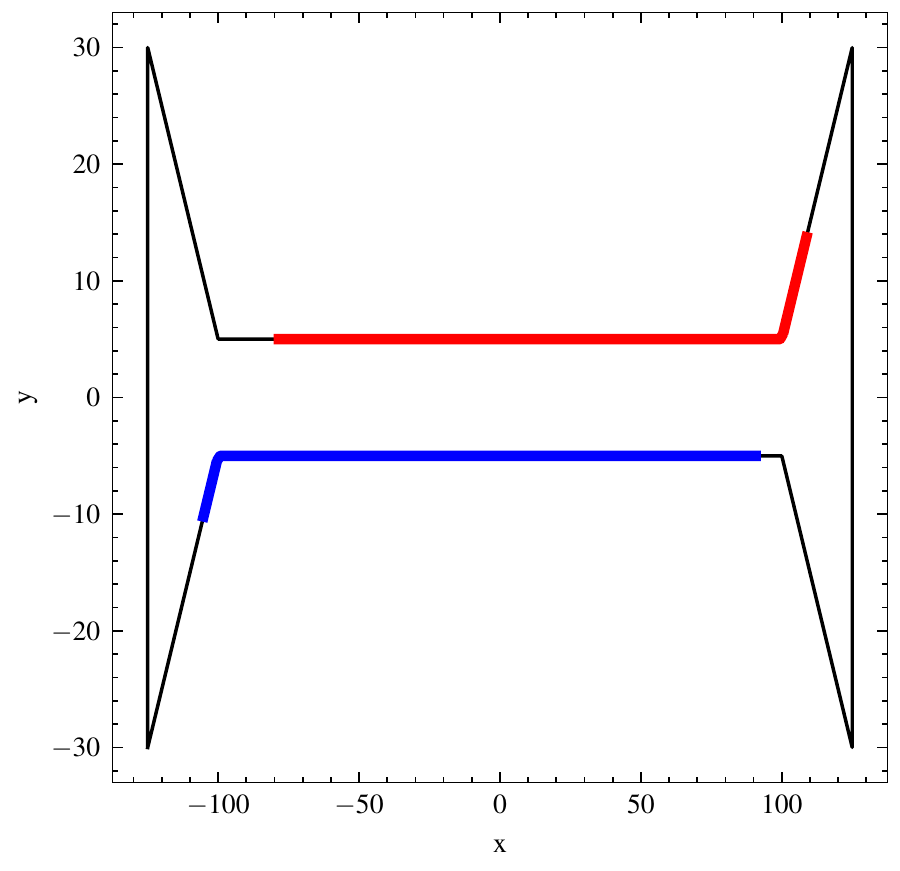}
	\end{subfigure}%
	~ 
	\begin{subfigure}[t]{0.25\linewidth}
		\centering
		\includegraphics[width=\linewidth]{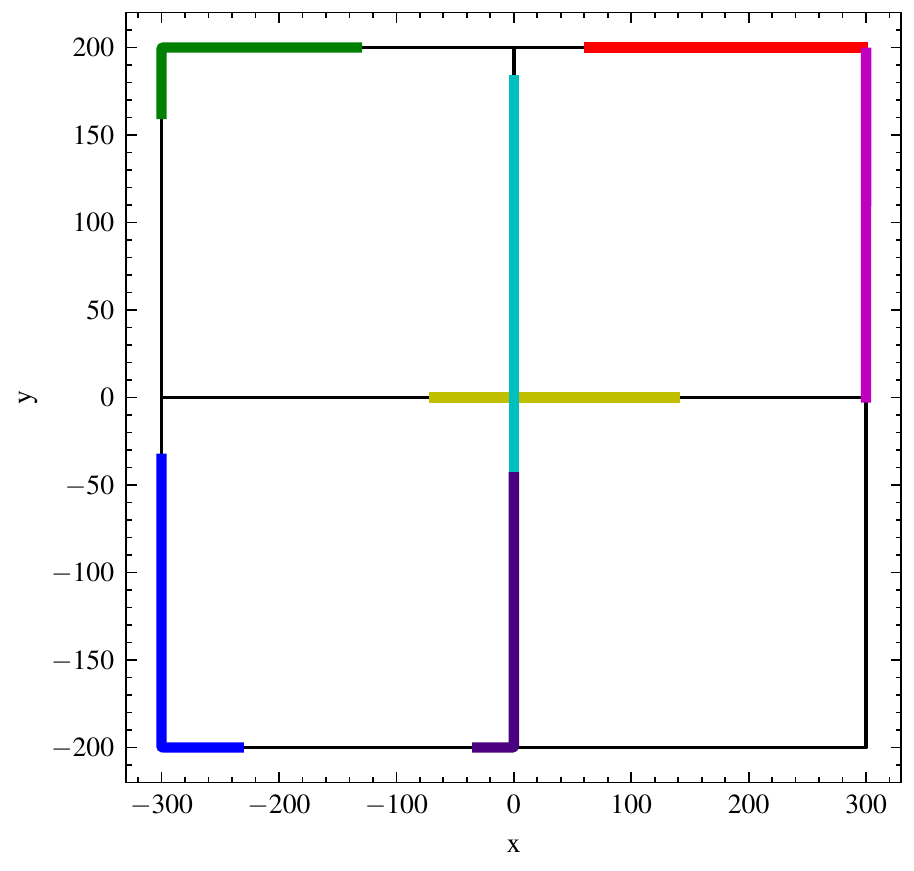}
	\end{subfigure}
	\caption{True target tracks (in different colors) for the three simulated scenarios, where black lines are road segments.  In scenario 1 (left), one target is initiated on the horizontal segment before the fork junction, which eventually leads to a track split. In scenario 2 (center), two targets are initiated well separated, before entering the two horizontal segments, potentially passing close to each other. In scenario 3 (right), three targets are initiated on random segments and more targets are born from different locations during the simulation.}
	\label{fig: scenario 1, 2, and 3}
\end{figure*}

In the simulation, sensors are mobile and independent with initial mean speed $\SI{12.3}{\metre/\second}$ (standard deviation is $\SI{1.5}{\metre/\second}$), and limited field of view $[-30,30]$ (\si{\metre}) in longitudinal direction.  The designed path for targets (\eg, a sidewalk) is assumed all observed within a sensor's field of view.  The parameters of the measurement model are 
\begin{align*}
    H
    &= \begin{bmatrix}
      1 & 0 \\
      0 & 1
    \end{bmatrix},
    &
    R
    &= \begin{bmatrix}
      \sigma^2  & 0 \\
      0  & \sigma^2/4
  \end{bmatrix}
\end{align*}
where $\sigma=\SI{0.5}{\metre}$. The clutter intensity is $\lambda_{\fa}(z) = \lambda V\mathcal{U}(z)$ where $\mathcal{U}(\cdot)$ is a uniform density over the sensor field of view, $V = 2 \times \SI{30}{\metre}$  is the network-constrained ``volume" of the field of view, and $\lambda = 0.01 \times  \SI{}{\metre}^{-1}$ is the average number of clutter observations per unit volume. Further, the discrete state information included in each provided observation is accurate.  In situations where several sensor platforms operate simultaneously, a sequential measurement update strategy is employed.  That is, a separate measurement update step is applied for each sensor.  This approach scales well with the number of sensors, and fits naturally in a distributed setting as measurements can be dealt with as they arrive.

The NC-MHT performance is compared to a standard MHT that describes the kinematic state by the target's position and velocity in two dimensions, and where the motion follows a constant velocity model.  Similarly, sensor observations are given in Cartesian $x$- and $y$-coordinates.

\textbf{Note:} For the simulation, target and clutter observations  were generated in the network-constrained space, yet used for both NC-MHT and standard MHT when comparing both performances.  This ensures using the same measurements when comparing both filters, however, it considerably benefits the two-dimensional standard MHT.  Also, the choice of linear models benefits the performance of the standard MHT, as compared to the NC-MHT.

\subsection{Variable Negative Information}
To save computations this implementation exploits the vital knowledge provided by so-called negative information, as described in Section \ref{sec: track hypotheses and track scores}. Regularly providing the filter with measurements of targets that should have been seen by the sensor, \ie, non-presence, has significant impact on track likelihood scores and can help abandon unlikely tracks.  Hence, the MTT should be updated with all scans, empty or not to maintain proper track scores.  It is tempting to drop empty scan, as they do not affect the track estimates just the scores, in order to save time.  To evaluate the importance of negative information and see its effect in the presented setting, a scheme with a varying proportion of empty scans reported to the MTT is proposed in the third scenario.

\subsection{Results}

Three simulated scenarios were used in this study. The first two are designed to  highlight the benefits of the proposed NC-MHT as compared to the standard MHT when it comes to target prediction and measurement association.  The third scenario evaluates the measurement dependency of the filter given a different number of observations.  For the first two scenarios $500$ Monte Carlo runs are performed, whereas the third, including two variations, is run $100$ times each. The presented results are averaged over the Monte Carlo runs.  For all three simulations, the maximum number of hypotheses per cluster and the minimum normalized hypothesis score are set to $50$ and $4$ respectively.  Also note that only tracks that have been assigned at least twice with a measurement have been kept for presentation in the global hypothesis. \\

\textbf{Scenario 1}
In the first scenario, one target initiated ahead of the fork junction was simulated for $100$ time steps.  True trajectories are shown in Fig.~\ref{fig: scenario 1, 2, and 3}.  For this scenario, $10$ sensor platforms were initiated randomly distributed in the network. The scenario parameters were set to $P_{\D}=0.95$, $\lambda_{\nt}=0$, $\lambda_{\fa}=0.6$.  

This scenario particularly highlights beneficial network constraint effects for more efficient target prediction over an extended period of time.  This scenario is challenging because when the target leaves the fork junction, and no measurement update is performed, correctly predicting the target state is complicated.  The network constraints allow to keep track of the two different turns the target may perform as it leaves the fork intersection, whereas the standard filter predicts the target to continue straight and going off-road eventually losing track, unless an observation is obtained right after the turn. This behaviour is illustrated in Fig.~\ref{figure: Illustration tracking behaviour}, where the standard MHT misses the turn in absence of new updates and continues its prediction in the wrong direction ($\tau^1_{1}$--$\tau^1_{5}$).  With the arrival of new measurements, the MHT reacts by initiating a new track with along the new road segment ($\tau^3_{6}$--$\tau^3_{7}$).

\begin{figure*}[!ht]
\minipage{0.48\textwidth}
  \includegraphics[width=\textwidth]{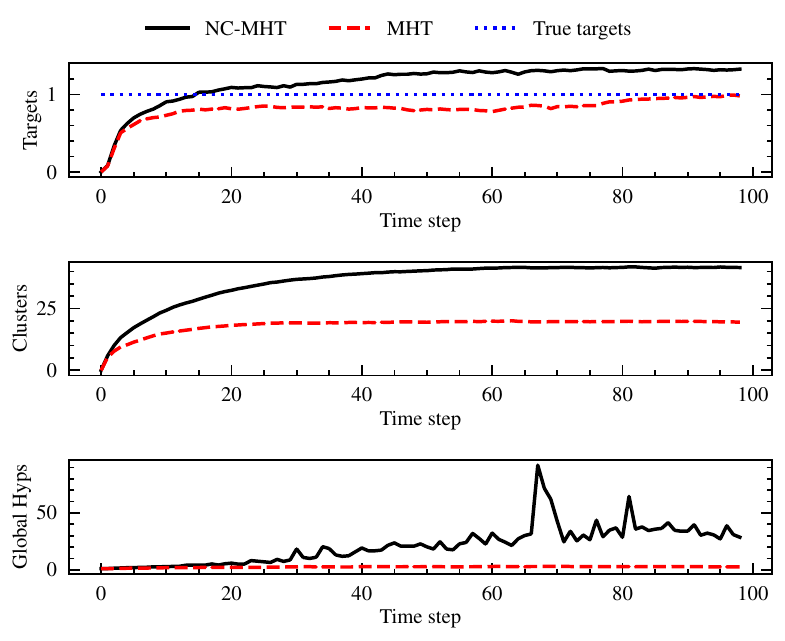}
    \caption{Statistics for simulation scenario 1.}\label{fig: stats scenario 1}
\endminipage\hfill%
\minipage{0.48\textwidth}
  \includegraphics[width=\textwidth]{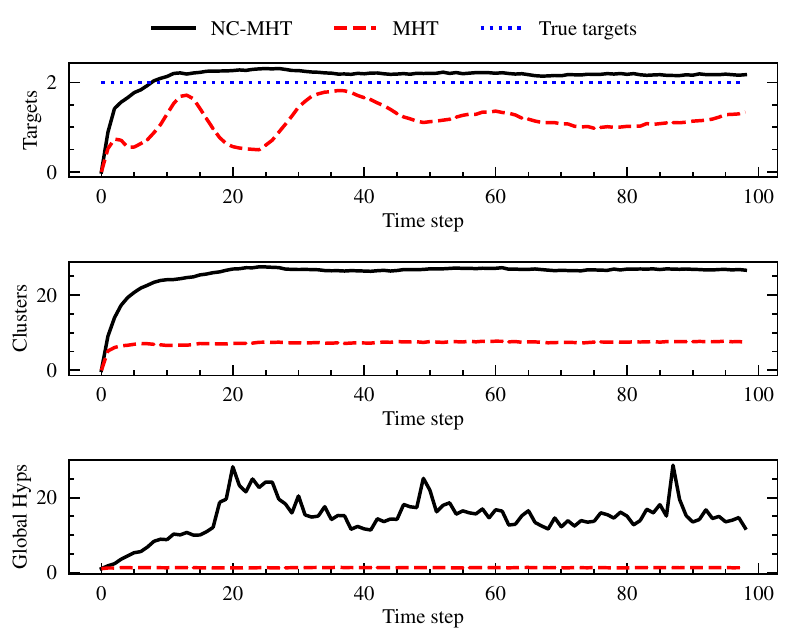}
    \caption{Statistics for simulation scenario 2.}\label{fig: stats scenario 2}
\endminipage
\end{figure*}

\begin{figure*}[!htb]
\minipage{0.48\textwidth}
  \includegraphics[width=\textwidth]{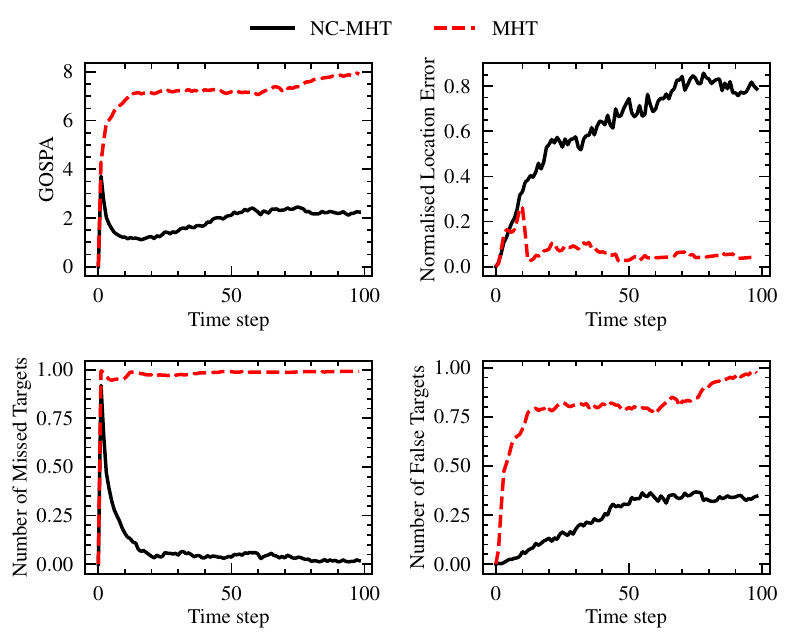}
    \caption{Tracking results for simulation scenario 1.}\label{fig: gospa scenario 1}
\endminipage\hfill%
\minipage{0.48\textwidth}
  \includegraphics[width=\textwidth]{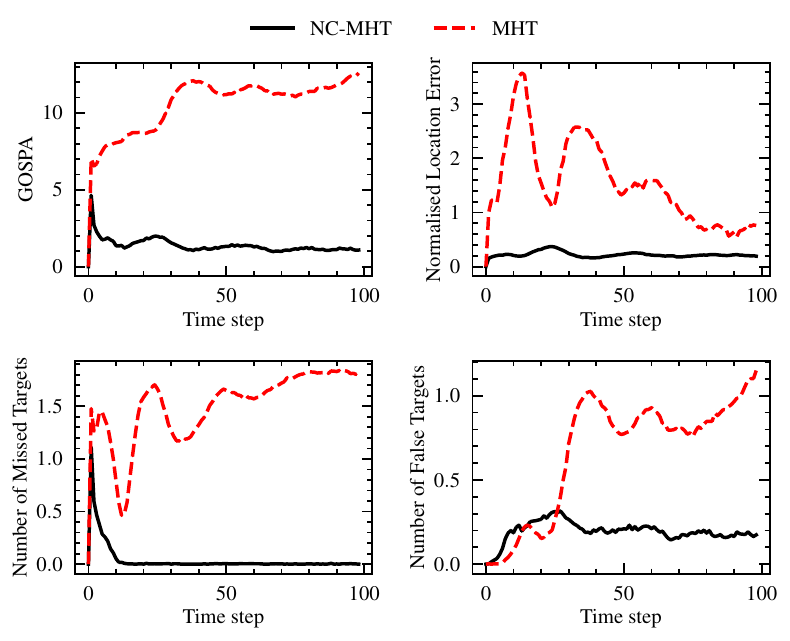}
    \caption{Tracking results for simulation scenario 2.}\label{fig: gospa scenario 2}
\endminipage
\end{figure*}

The GOSPA performance is shown in Fig.~\ref{fig: gospa scenario 1}, and summarizing statistics are shown in Fig.~\ref{fig: stats scenario 1}.  A closer look at the latter unveils a higher number of tracks produced by the NC-MHT, here resulting in a higher number of clusters and global hypotheses.  This a likely bi-product of the clutter model benefiting the standard MHT and can be improved by adjusting tuning parameters.  Overall, for this scenario the GOSPA results show that the NC-MHT gives smaller errors than the standard MHT filter.  The GOSPA location error is larger for the NC-MCT towards the end, this is a result of the NC-MHT being better at maintaining tracks, whereas the MHT lose them effectively moving part of the location error to a missed track error.  This is further substantiated by Table~\ref{tab: results scenarios 1-3}.  Hence, in cases where maintaining track continuity is of importance the proposed NC-MHT has benefits. \\

\textbf{Scenario 2}
In the second scenario, two targets are initiated well separated before moving onto a horizontal segment in opposite direction were simulated for $100$ time steps.  True trajectories are shown in Fig.~\ref{fig: scenario 1, 2, and 3}.  For this scenario, $20$ sensor platforms were initiated randomly distributed in the network. The scenario parameters were set to $P_{\D}=0.95$, $\lambda_{\nt}=0$, $\lambda_{\fa}=0.6$.

This setup highlights how adding network constraints simplifies the measurement  association process. This scenario is challenging for traditional approaches because when the targets are close their measurements form a single cluster, making data association difficult.  Given the type of targets, their speeds are assumed low and thus do not allow to distinguish their movements in two different directions that way.  Fig.~\ref{figure: ncmhtvsmht} illustrates how traditional approaches without network restriction need to consider all incoming measurements (if in the gate) for the measurement to track association, and possibly fail.  The network constraints allow to bypass growing free-space complexity of traditional approaches and lead to more efficient gating and data association as measurements as motion is bound to a specific segment. 

The GOSPA performance is shown in Fig.~\ref{fig: gospa scenario 2}, and summarizing statistics are shown in Fig.~\ref{fig: stats scenario 2}.  Overall, for this scenario the GOSPA results show that the NC-MHT performs better than the standard filter. The NC-MHT can keep tracks longer in time, as shown in Table~\ref{tab: results scenarios 1-3}.  Note that, as compared to the first scenario, Scenario 2 has twice the number of sensors, hence resulting in more observations.  This could explain the lower location error.  As expected, results suggest that NC-MHT number of misses significantly less targets. \\

\textbf{Scenario 3}
In the third scenario, three targets initiated at random locations in the network are simulated for $100$ time steps.  New targets enter in the surveillance area at different time steps and are born at random locations in the network.  Targets leaving the surveillance area is not considered here, however, can be easily handled by adjusting the probability of survival $P_S$ for future implementations.  The true trajectories are shown in Fig.~\ref{fig: scenario 1, 2, and 3}.  For this scenario, the performance of the filters were compared based on two realisations with different numbers of: a) number of available sensors, and b) proportion of empty scans provided to the tracker. The scenario parameters are shown in Table~\ref{tab: parameters scenario 3}.  

This scenario illustrates a slightly more realistic urban traffic setting and is challenging because of the many intersections, targets, and observations. This simulated scenario particularly highlights the potential of the approach for real-world applications. More precisely, Scenario  3a) highlights the (low) density of observations needed as a result of efficient predictions along the road segments in the proposed method.  Scenario 3b), on the other hand, explores the importance of the negative information provided to the tracker.

\begin{figure}[htb!]
\centering
  \includegraphics[width=\columnwidth]{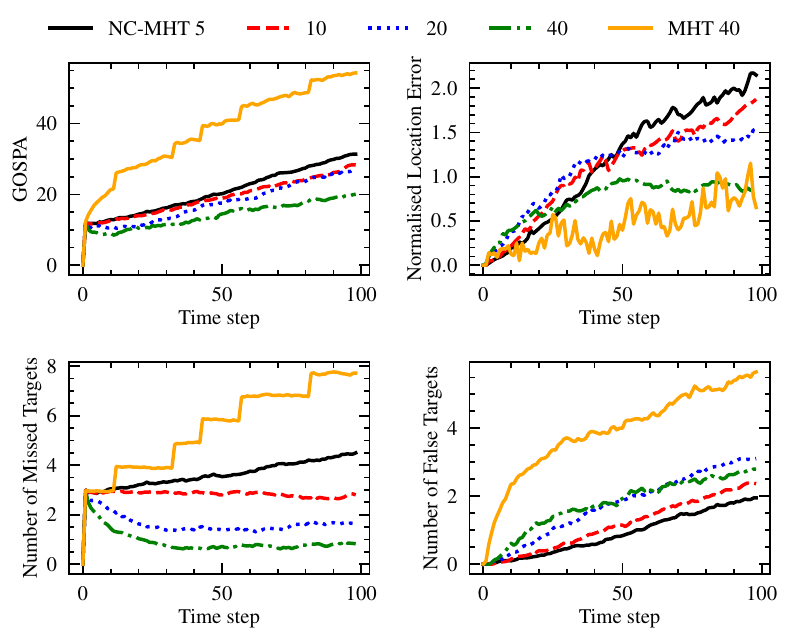}
    \caption{Tracking results for simulation Scenario 3a). The legend refers the number of sensors initiated.}\label{fig: gospa scenario 3 (vcls)}
\end{figure}

Fig.~\ref{fig: gospa scenario 3 (vcls)} shows the GOSPA performance of the NC-MHT filter given $5, 10, 20$ and $40$ mobile spatially distributed sensor platforms respectively, and the performance of the MHT with $40$ initiated sensors.  Results show that the NC-MHT achieves a similar performance to the standard filter yet with significantly less sensors.  In this simulation, the lower the number of sensors, the fewer measurements are provided to the tracker, and the less updates can be performed by the filter.  This confirms results from scenario~1 suggesting that the included network structure knowledge of the NC-MHT allow for more efficient target predictions over extend period of time.  The difference in track length shown in Table \ref{tab: results scenarios 1-3} supports this outcome.  More sensors means more clutter measurements, but also more observations of true targets.  Results also show that with more sensors, more unlikely tracks can be killed, \eg, existing tracks get less likely when sensors expected a target but did not see any.

\begin{figure}[htb!]
\centering
  \includegraphics[width=\columnwidth]{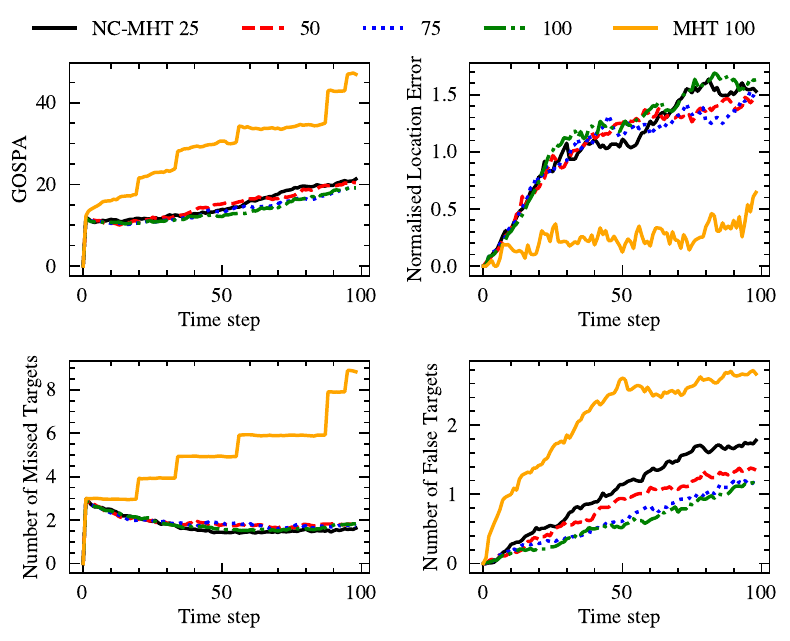}
    \caption{Tracking results for simulation scenario 3b). The legend refers to available proportion of available sensors (in percent) that provide the tracker with additional negative information, respectively.}\label{fig: gospa scenario 3 (empty)}
\end{figure}

Fig.~\ref{fig: gospa scenario 3 (empty)} shows the GOSPA performance of the NC-MHT filter given $25$, $50$, $75$, and $100$ percent of available sensors that provide the tracker with additional negative information, respectively, and is compared to the performance of the MHT given all sensors are included in the time update step.  The results indicates that adding only a few instances of negative information can benefit the tracker significantly.

\begin{table}[t!]
    \caption{Parameters for the two realisations of simulation Scenario 3.}
    \label{tab: parameters scenario 3}

    \centering
    \begin{tabular}{lccc|cc}
    \hline
         & \boldmath $P_{\D}$ & \boldmath$\lambda_{\nt}$ & \boldmath$\lambda_{\fa}$ & \bfseries sensors & \bfseries empty scans \\ \hline 
         S3a) & $0.95$ & $2.4$ & $0.6$ & $5, 10, 20, 40$ & $1$ \\
    S3b) & $0.95$ & $2.4$ & $0.6$ & $20$ & $25\%, 50\%, 75\%, 100\%$ \\
    \hline
    \end{tabular}
\end{table}

\begin{table*}[t!] %
    \caption{Results for all Scenarios 1--3.}
    \label{tab: results scenarios 1-3}

    \centering %
    \begin{tabular}{lcccccccc|cc}
    \hline
         & \multicolumn{2}{c}{\bfseries GOSPA} & \multicolumn{2}{c}{\bfseries NLE} & \multicolumn{2}{c}{\bfseries MT} & \multicolumn{2}{c}{\bfseries FT} & \multicolumn{2}{|c}{\bfseries Track length} \\
            & NC-MHT & MHT & NC-MHT & MHT & NC-MHT & MHT & NC-MHT & MHT & NC-MHT & MHT \\ \hline 
         S1 & $411.7$  & $848.3$ & $224.5$ & $199.3$ & $217.5$ & $2838.8$ & $2396.9$ & $4732.0$ & $52.8$ & $26.9$ \\
         
         S2 & $159.2$  & $860.0$ & $37.5$ & $1086.2$ & $52.5$ & $4532.6$ & $708.7$ & $2264.0$ & $10.3$ & $5.7$ \\
         
         S3.a - 5 & $1095.5$  & n.a. & $1313.1$ & n.a. & $9192.3$ & n.a. & $2477.4$ & n.a. & $27.1$ & n.a. \\
         \hphantom{S3.a }- 10 & $1004.5$  & n.a. & $1611.6$ & n.a. & $6389.7$ & n.a. & $3072.9$ & n.a. & $33.0$ & n.a. \\
         \hphantom{S3.a }- 20 & $857.3$ & n.a. & $1404.3$ & n.a. & $3927.0$ & n.a. & $3127.3$ & n.a. & $36.4$ & n.a. \\
         \hphantom{S3.a }- 40 & $693.7$  & $1482.5$ & $894.6$ & $198.8$ & $1890.8$ & $9758.7$ & $3182.0$ & $13027.5$ & $39.2$ & $37.1$ \\
         
         S3.b - 25 & $1055.1$  & n.a. & $1467.5$ & n.a. & $3787.2$ & n.a. & $7058.2$ & n.a. & $58.1$ & n.a. \\
         \hphantom{S3.b }- 50 & $944.1$  & n.a. & $1545.1$ & n.a. & $3693.1$ & n.a. & $4827.8$ & n.a. & $49.0$ & n.a. \\
         \hphantom{S3.b }- 75 & $897.5$  & n.a. & $1556.5$ & n.a. & $3836.8$ & n.a. & $3778.8$ & n.a. & $42.2$ & n.a. \\
         \hphantom{S3.b }- 100 & $878.5$  & $1485.4$ & $1425.3$ & $185.9$ & $3931.8$ & $11152.6$ & $3421.7$ & $11883.2$ & $37.8$ & $32.0$ \\
    \hline
    \end{tabular}
\end{table*}

\section{Outlook} \label{section: Outlook}
In this section, we discuss technical and non-technical effects that may be induced by design decisions on the tracker, and the tracking system.

\subsection{Motion and Target Dynamics}
A comprehensive understanding of targets motion and their respective environment is an important factor for better trajectory reconstruction.  Extending the proposed framework using more sophisticated kinematic models can easily be achieved, and could lead to better prediction performances on the short-term.  As standard models are widely accepted for their simplicity, it seems important to keep their accuracy for trajectory estimation in balance with inherent computational cost.  Yet, the longer the prediction horizon needs to be, the more important choice of the model becomes.  Moreover, it is about how to make better predictions when little information is available. On the other hand, improving the scene understanding using social- and behaviour-aware models could significantly strengthen tracking and resulting trajectory estimations (\eg, \cite{Krishanth_2017}), for instance, in shared spaces, when approaching signalized intersections, or during overtaking maneuvers.  This additional complexity however is a non-trivial extension to our proposed MHT approach as it breaks the assumption of independent targets.  However with some approximations, \eg, that the interactions do not interfere with anything else than the propagation of targets, these models could probably be used with good results and at a reasonable cost.  With pedestrian and cyclist motion being influenced by weather, time of day, surrounding infrastructure and environment, as well as social and behavioural cues, future work would benefit from including such contextual information as this can lead to better long-term, and potentially network-wide, tracking results. Future extensions of this work could also substitute fixed parameters with learned distributions, which helps the tracker to better interpret measurements and tracks (\eg, \cite{Luber_2011}). For instance, the rate at which new targets appear and where they appear are dependent on time and space and could similarly be learned.

\subsection{Sensing Infrastructures and Measurements} \label{subsection: sensing infrastructure and measurements}
The mobile sensing setting used in this work can conceptually be extended to any type of sensing realm where incoming measurements provide information about targets, and regularly report non-presence to improve tracking. Also since current developments hint to the integration of additional decentralized processing layers where computing power is brought to the edge, thereby fusion of additional information from additional static or participatory sensor infrastructure is conceivable and represents no extra adaptation to the proposed method.

Contextual information about the static and dynamic environment can be captured by different types of sensors. Using additional information could lead to significant improvements of the tracking system, and particularly in the data association logic, for instance as a way to strengthen the discrimination between different objects.  Recent advances in deep-learning-based trackers, for instance, improve the discriminative power for visual tracking applications \cite{Zhang_improved_2021}.

Furthermore, node transition probabilities can either be determined based on prior knowledge or set based on empiric studies about, \eg, how many take a left or right turn at an intersection.  Following traffic patterns this could be something that changes throughout the day and week and could be function of environment features (\eg, weather, green canopy), real-time incidents (\eg, events, construction), network knowledge (\eg, road types, touristic routes) or infrastructure (\eg, road quality). 

\subsection{Privacy Considerations}
Further improvements of the NC-MHT tracking performance can make it more suited for real-world applications, yet the gain in utility may come at the cost of the targets' privacy.  The centralised scheme presented in this work, which pools all location observations to a central entity and hence may facilitate the surveillance of individuals or collectives, raises privacy concerns.  

Individual location traces are akin fingerprints, as demonstrated by past research on the similarity of human mobility patterns \cite{Alessandretti_2020}, their predictability \cite{Song_2010}, and the uniqueness of individual traces \cite{De_montjoye_2013}.  But it is the realisation that with sufficient data it is possible to precisely locate, track, and possibly (re-)identify people that is a major source of concern.  Standard MTT approaches have been frequently used in the location privacy community, \eg, Gruteser and Hoh were among the first to use a MHT to link anonymized location samples to some individual users and recreate their traces \cite{Gruteser_2005}. 

Anonymized location samples as considered in this paper, however, create a false sense of security because of the the spatial and temporal correlation between successive observations.  Longer tracking duration, for instance, typically leads to an accumulation of information and can help infer sensitive information about individuals.  The traffic density, on the other hand, can have both positive and negative impact on privacy, but is a factor beyond an individual's control.  Using auxiliary data sources (\eg, household address or location-based service databases) to the anonymous traces further increases the risk of inferring locations, identities, and potentially interests of an individual.  

Overall, the accumulation of observations in a central database, anonymized or not, reduces the burden for attacks that can easily be automated and applied to large numbers of individuals, therefore augmenting the risk beyond individual tracking to a more collective threat.  

More work is needed to study effects of the methodological, and higher-level system design space, both on target privacy and utility of future applications.
 
\section{Conclusion} \label{section: conclusion}
The advent of mobile sensor platforms is expected to drastically increase the amount of information collected about traffic participants (\eg, pedestrians or cyclists) in populated environments, which can be of great value for multiple future applications, e.g., digital twin, smart road infrastructure, autonomous driving, and advanced traffic surveillance.  To make use of this information, this work presented a framework for network-constrained tracking of targets with observations from spatially distributed and mobile sensor platforms. The key contribution represents the introduction of network bound targets into the multi-target tracking problem.

The proposed approach can be applied to any type of network-constraint environment, bypassing the growing free-space complexity of traditional MTT approaches, and use different types of observations. The generic nature of this work makes it interesting for any type of sensing setting where incoming observations provide information about targets, and potentially report non-presence of targets.  Derived information can, for instance, be valuable input for traffic signal controller aiming at reducing idling times for cyclists at signalized intersections, for crowd monitoring systems measuring previously unobserved network links, or fusing observations from different sensors and road users to distribute traffic at the micro level.  Overall, we see the potential to utilize this information in more places as it can be applied to more difficult problems than those addressed before. 

Future work will explore the application space for next generation traffic surveillance and control systems, evaluating a range of simulated scenarios using a microscopic traffic simulator.  The highlighted properties of the modified method as compared to a standard MHT can be used to predict the gains obtainable with other MTT filters.  A more thorough evaluation, then focused on a particular scenario or application is an interesting direction for future research.  Furthermore we motivate the need for experiments in real road traffic scenarios, which can help increase the persuasiveness of the proposed framework.  It should be noted that given the (at this time) difficult access to data from a fleet of moving sensor platforms over longer time and space, any other dataset that provides information about targets and regularly report non-presence could be applied to our framework.  This work hints to future directions for research in tracking and privacy engineering within the intelligent transportation realm.

\IEEEpeerreviewmaketitle

\section*{Acknowledgment}
This work was supported by the European Research Council and Amsterdam Institute for Advanced Metropolitan Solutions through the ALLEGRO project under Grant 669792.  G. Hendeby has received funding from the Center for Industrial Information Technology at Link\"oping University
\mbox{(CENIIT) Grant 17.12.}

\bibliographystyle{IEEEtran}
\bibliography{bibliography.bib}	


\end{document}